\documentclass[twocolumn,floatfix,aps,showpacs,showkeys]{revtex4-1}
\usepackage{epsfig,subfigure,latexsym}

\usepackage{amssymb,amsmath}

\begin{document}

\title{Boson star at finite temperature}

\author{S. Latifah, A. Sulaksono, and T. Mart}

\affiliation{Departemen Fisika, FMIPA, Universitas Indonesia, Depok 16424, Indonesia }

\begin{abstract}
By using a simple  thermodynamical method we confirm the finding of Chavanis and Harko that stable Bose-Einstein condensate stars can form. However, by using a thermodynamically consistent boson equation of state, we obtain a less massive Bose-Einstein condensate star compared to the one predicted by Chavanis and Harko. We also obtain that the maximum mass of a boson star is insensitive to the change of matter temperature. However, the mass of boson star with relatively large radius depends significantly on the temperature of the boson matter.
\end{abstract} 


\maketitle
Since the Bose-Einstein condensation has been observed and well studied in laboratory, the possible existence of this object and its implications on astrophysical or cosmic scales have recently attracted much attention. The consequences if dark matter were in the form of Bose-Einstein condensation and finite temperature boson matter have been extensively studied (see e.g., Ref.~\cite{HM2012} and the references therein).  It is shown in Ref.~\cite{Guzman2006} that self-gravitating Bose-Einstein condensates may form a gravitationally stable structure. This is the basis of the boson star formation. The  degenerate or non-degenerate boson stars,  have been also investigated by many authors  not only by using relativistic but also non-relativistic approaches as well as by employing different techniques~\cite{chavanis,Bilic2000,Colpietc}. For the latest review on boson star we refer the reader to Refs.~\cite{LP2012,SM2003}.

Recently, Chavanis and Harko~\cite{chavanis} have studied the Bose-Einstein condensate by using the Gross-Pitaevskii equation with an arbitrary non-linearity. To formulate the dynamics of the system, they used the continuity and hydrodynamics Euler equations. In the case of a condensate with quartic non-linearity, they found a polytropic equation of state (EOS) with index one. They also found that the condensates with particle masses of the order of two neutrons mass and scattering length of the order of 10-20 fm produce a maximum mass of non-rotating star-like object of the order of $2M_\odot$, whereas the corresponding radius is in the range of 10-20 km. {Based on this result they considered that the recently observed pulsar~\cite{Demorest} with the mass of around  $2M_\odot$ might be connected to the Bose-Einstein condensate star~\cite{chavanis}. However, it should be noted that in Ref.~\cite{chavanis} and in this paper only the non-rotating solutions are investigated. Furthermore, there exist enough successful models for explaining massive pulsars in the literature (see for examples Refs~\cite{MCM2013,CHZF2013,LP2010}).}

In this Brief Report we confirm the finding of Ref.~\cite{chavanis} by using a different method to calculate the EOS and extend the calculation to the case of boson star at finite temperature. We use thermodynamical method and with a simple mean-field approximation in Hamiltonian density.

The Hamiltonian operator of a non-relativistic interacting identical bosons system ~\cite{cjpethick,chavanis} with free boson mass $m$ as the zero energy reference can be written as
\begin{eqnarray}
\hat{H} & = & \int d {\mathbf r} \hat{\Psi}^{\dagger} ({\mathbf r}) \left[ -\frac{1}{2m}~\nabla^2 + m \right] \hat{\Psi} ({\mathbf r}) \nonumber\\
&+& \frac{1}{2} \int d {\mathbf r} d {\mathbf r}' \hat{\Psi}^\dagger ({\mathbf r}) \hat{\Psi}^\dagger ({\mathbf r}') V ({\mathbf r} - {\mathbf r}') \hat{\Psi} ({\mathbf r})\hat{\Psi} ({\mathbf r}'),
\label{hamilton1}
\end{eqnarray}
where $\hat{\Psi} ({\mathbf r})$ and $\hat{\Psi}^\dagger ({\mathbf r})$ are the boson field annihilation and creation operators at the position ${\mathbf r}$, while  $V ({\mathbf r} - {\mathbf r}')$ is the general two-body potential. The  additional free boson mass term in Eq.~(\ref{hamilton1}) is important to obtain a correct energy density ($\varepsilon$) of the bosons system. Note that we use $\hbar=c=k_B=1$ as a  convention. 

\begin{figure}
\epsfig{figure=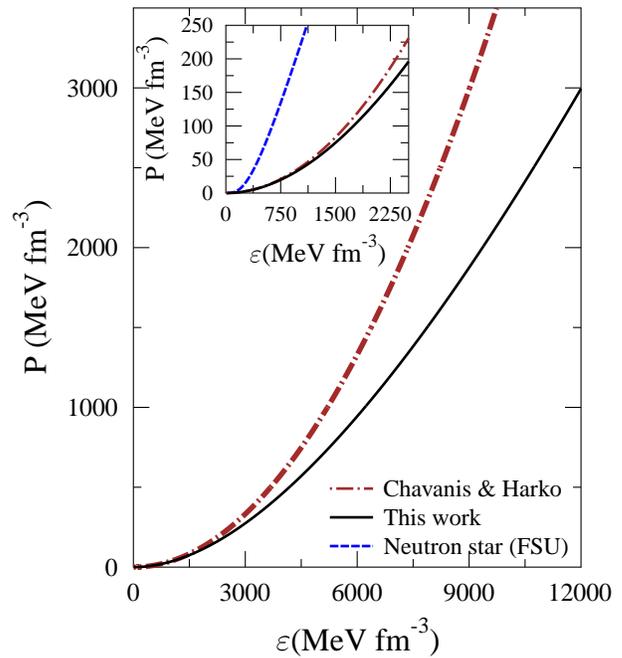, width=8cm}
\caption{(Color online) The equation of state (EOS) of boson matter in the case of { $T$ $\to$ 0} used as input to obtain the mass-radius plots with (solid line) and without (dash-dotted line) the  $\frac{1}{2} u_0 n_0^2$ term in the energy density. The inserted panel exhibits the corresponding EOS at low pressure.}
\label{fig_EOST0}
\end{figure}

In the case that the interaction between two bosons is short range, the two-body potential may be approximated as a contact interaction~\cite{chavanis}, i.e., $V ({\mathbf r} - {\mathbf r}') \to u_{0} \delta ({\mathbf r} - {\mathbf r}')$, where { the coupling parameter  $u_{0}$ indicates the strength of the two body interaction.  The physical motivation of this approximation is that in dilute and cold boson gas only binary collisions at low energies are relevant, whereas these collisions are characterized by a single parameter, i.e., the $s$-wave scattering length $a$, independent of the details of the two body potential. Therefore, it can be shown that $u_{0}=4\pi a/m$ (see, e.g., Refs.~\cite{cjpethick,chavanis} for further explanation)}. If we substitute the boson field in Eq.~(\ref{hamilton1}) with its explicit form, i.e., $\hat{\Psi} ({\mathbf r}) = \sum_{\alpha} \Psi_{\alpha} ({\mathbf r}) \hat{\bf a}_{\alpha}$, where  $\hat{\bf a}_{\alpha}$ is the boson annihilation operator, and rewrite the Hamiltonian in momentum basis, we obtain the Hamiltonian density in the form of  
\begin{eqnarray}
\hat{\mathcal{H}} & = & \sum_{k}  \frac {k^2}{2m}\hat{n}_{k} + \left[ \sum_{k} m  \hat{n}_{k} + \frac{1}{2} u_{0} \sum_{k,k'} \hat{n}_{k} \hat{n}_{k'}\right]\nonumber\\ & \equiv & \hat{\mathcal{H}}_f + \hat{\mathcal{H}}_i .
\label{hamiltondensity3}
\end{eqnarray}  
In Eq.~(\ref{hamiltondensity3}) $\hat{n}_{k}$ is the occupation number operator in state $k$. When the number of bosons involved becomes very large (dense matter) and the temperature ($T$) is not too high (much less than the boson mass), the role of the quantum fluctuation that enters into the energy density becomes less significant compared to that of the nonlinear interactions~\cite{chavanis}. In this situation, we may approximate the interaction Hamiltonian density with its expectation value,  i.e., $\hat{\mathcal{H}}_i$ $\approx$ $<\hat{\mathcal{H}}_i>$$\equiv$ ${\mathcal{H}}_i$. At first glance this approximation seems like a draw back. However, by using this approximation, we obtain a thermodynamically consistent relation of the pressure and energy density of an interacting boson system for finite temperature and the result can be expressed in a simple analytical form. Therefore,  due to the latter, it can be compared easily to other analytical results such as the one predicted by  Thomas Fermi approach~\cite{chavanis}. In the limit of $T$ $\to$ 0, it can be seen later that the pressure of  boson system predicted by this approximation is consistent with the one obtained by using Thomas Fermi approach~\cite{chavanis}. Furthermore, this result can be used as a benchmark for further investigation by using similar method, but with a more refined numerical approximation. To this end, we can obtain an approximate form of the grand canonical partition function of Eq.~(\ref{hamiltondensity3}) as
\begin{equation}
\mathcal{Z}_B\approx e^{-\beta V(\frac{1}{2} u_0 n^2-m n)}
\mathcal{Z}_f,
\label{partfunc}
\end{equation}
where $\beta$ = ${1}/{T}$, $V$ is the volume of the boson system, and $n$ = $<\hat{n}>$ = $\sum_k <\hat{n}_k>$, while $\mathcal{Z}_f$ is the partition function for free bosons. By using grand canonical potential density, $\tilde{\Omega}_{B}=-\ln\mathcal{Z}_B/{V \beta}$, we can calculate the pressure, energy density and number density of interacting boson system with the density dependent interaction through~\cite{yin}
\begin{eqnarray}
P &=& -\tilde{\Omega}_{B} + n{\left(\frac{\partial \tilde{\Omega}_B}{\partial n }\right)}_{T,\mu}\nonumber\\
\varepsilon &=& \tilde{\Omega}_{B} +\mu n -T { \left(
\frac{\partial \tilde{\Omega}_{B}}{\partial T} \right)}_{\mu,n}\nonumber\\ 
n&=&- {\left( \frac{\partial\tilde{\Omega}_{B}}{\partial\mu} \right)}_{T,V},
\label{EOS}
\end{eqnarray}
where $\mu$ is the chemical potential of the system. It can be proved easily that the pressure and energy density of the system in Eq.~(\ref{EOS}) is thermodynamically consistent because they obey the relation~ $P$ = $n^2 {\{ {\partial (\varepsilon/n)}/{\partial n} \}}_T$ \cite{glendenning}. In thermodynamical limit \cite{greiner}, i.e., $\sum_{k}\to \{{V}/{(2\pi)^3}\}\int d{\bf k}$, the pressure in Eq.~(\ref{EOS}) becomes
\begin{equation}
P = \frac{1}{2} u_0 n^2 + \frac{1}{\beta} \left[ \frac{1}{2^3} \left( \frac{2 m}{\pi \beta} \right)^{3/2} g_{5/2} (z) \right],
\label{Press}
\end{equation}
where the fugacity $z$ $\equiv$ $e^{\beta \mu}$ and $g_n(z)$=$\sum_{k=1}^{\infty} {z^k}/{k^n}$ for $0 \le z\le 1$. It is obvious that if  $u_0=0$ then the pressure reduces to the case of the non-degenerate free bosons system~\cite{greiner}, while in the limit of $T\to 0$ it becomes $P = \frac{1}{2} u_0 n_0^2$. This result is identical to the finding of Ref~\cite{chavanis}. In other words, in the limit of $T$ $\to$ 0 the prediction of this approximation is quite good. Therefore, it is reasonable to believe that for a system with nonzero $T$, but not too far from zero, a more refined approximation in the Hamiltonian density may not lead to a much different result and the trend remains the same.  Note that $n_0$ is the number density at ground state. On the other hand, the energy density reads
\begin{equation}
\varepsilon = m n + \frac{1}{2} u_0 n^2 + \mu n,
\label{edense3}
 \end{equation}
whereas in the limit of $T$ $\to$ 0 it becomes $\varepsilon$ = $m n_0 + \frac{1}{2} u_0 n_0^2$. The latter is in contrast to the energy density used in Ref~\cite{chavanis}, in which the second term is missing, in spite of the fact that this term is crucial for maintaining the thermodynamical consistency of  the pressure and energy density. Therefore, it is interesting to check how significant the role of the term $\frac{1}{2} u_0 n_0^2$ inside the energy density of the system in the boson star properties. In Fig.~\ref{fig_EOST0} we plot the EOS of boson matter with and without this term in the limit of  $T$ $\to$ 0. 
It is obvious that at low density or low pressure this term provides a negligible effect, but at high densities the effect of this term is quite significant, i.e., it stiffens the EOS. { For completeness,  we also provide the EOS of neutron star without the presence of exotics, such as hyperons, calculated by using the Florida State University (FSU) parameter set~\cite{FSU}, in the inset of Fig.~\ref{fig_EOST0}. Obviously, the EOS of boson matter has the same order of magnitude, but softer than that of the neutron star. However, by increasing the scattering length $a$ (the strength of interaction), we can obtain a stiffer boson star EOS. } We need also to note that the number density of boson matter is 
\begin{equation}
n = \left[ \frac{1}{2^3} \left( \frac{2 m}{\pi \beta} \right)^{3/2} g_{3/2} (z)  \right] + n_0.
\label{nden}
 \end{equation}
Obviously, the number density of the boson system in this approximation is similar to the one of free boson system and in the limit of  $T$ $\to$ 0 it becomes $n=n_0$. 

Using Eqs.~(\ref{Press}),~(\ref{edense3}), and ~(\ref{nden}) as the boson EOS input, we can obtain the structure of boson star by integrating the standard Tolman-Oppenheimer-Volkoff (TOV) equation~\cite{glendenning,saphiro}
 \begin{equation}
\frac{d P}{d r} = \frac{G (\varepsilon+P)(M+4 \pi r^3 P) }{r (r-2 G M)} ~~~, ~~~ 
\frac{d M}{d r} = 4 \pi \varepsilon r^2.
\end{equation}

\begin{figure}
\epsfig{figure=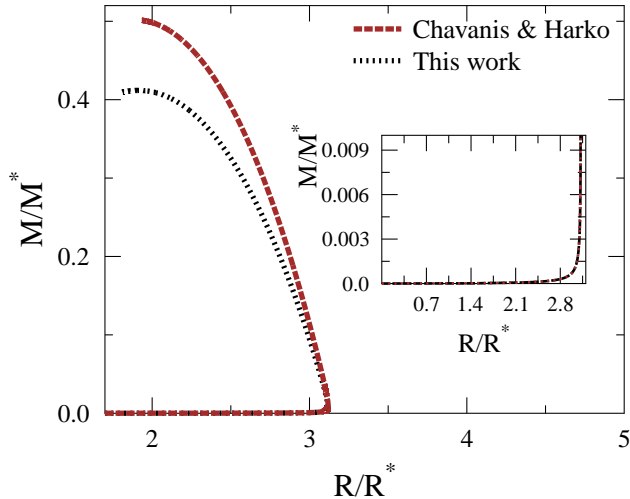, width=8.5cm}
\caption{(Color online) The mass-radius relation for a boson star at $T=0$. Solid line represents the result obtained by using the EOS of Eqs.~(\ref{Press}) and (\ref{edense3}), while the dash-dotted line is the result obtained by the authors of Ref.~\cite{chavanis}.}
\label{fig_RMT0}
\end{figure}
\begin{figure}
\epsfig{figure=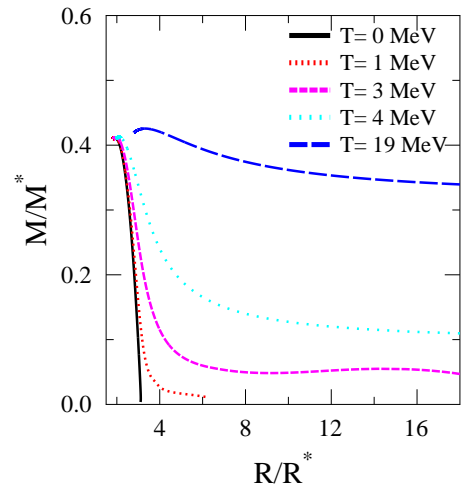, width=6cm}
\caption{(Color online) The mass-radius relation of a boson star for a reasonable range of temperature variation.  }
\label{fig_RMFT}
\end{figure}

Here, we consider that the interior of the star is composed in whole of boson matter without crust on the star surface. At the center of the star $M(0)=0$, while on the surface $P(R)=0$. In Fig.~\ref{fig_RMT0}, we show the boson star mass-radius relation for $T=0$ by taking into account the $\frac{1}{2} u_0 n_0^2$ term in the energy density and compare it with the case without this term, whereas in Fig.~\ref{fig_RMFT} we show the effect of temperature variation on  the structure of boson star. Note that in Figs.~\ref{fig_RMT0} and ~\ref{fig_RMFT}, we use the same definitions of $R^*$ and $M^*$ as the ones used in Ref.~\cite{chavanis}, i.e., $R^* = {\left({a}/{Gm^3}\right)}^{1/2}$ and $M^* = {\sqrt{a}}/{{Gm}^{3/2}}$. The inset in Fig.~\ref{fig_RMT0} enlarges the region with small value of $M/M^*$,  where we can see that both curves coincide with each other.  In our calculation, as for example, we take $a = 1$ fm, $m=2 m_n$, and $m_n$ is nucleon mass, {so that $R^*$ = 2.106 km and $M^*= 1.420~M_\odot$}. Definitely the mass and radius of the star depend quite sensitively on the values of $a$ and  $m$~\cite{chavanis}. {Figure~\ref{fig_RMT0} displays that by increasing the mass the radius increases up to the critical radius $R \approx 3.1 R^*$. After that the radius decreases as the mass increases up to the maximum mass. At this point the radius is minimum. Such a trend is similar to the case of the quark star~\cite{Li2010}. However, in the quark star mass-radius relation, in contrast to the case of boson star, the critical radius appears at a relatively higher mass}.  

\begin{figure}
\epsfig{figure=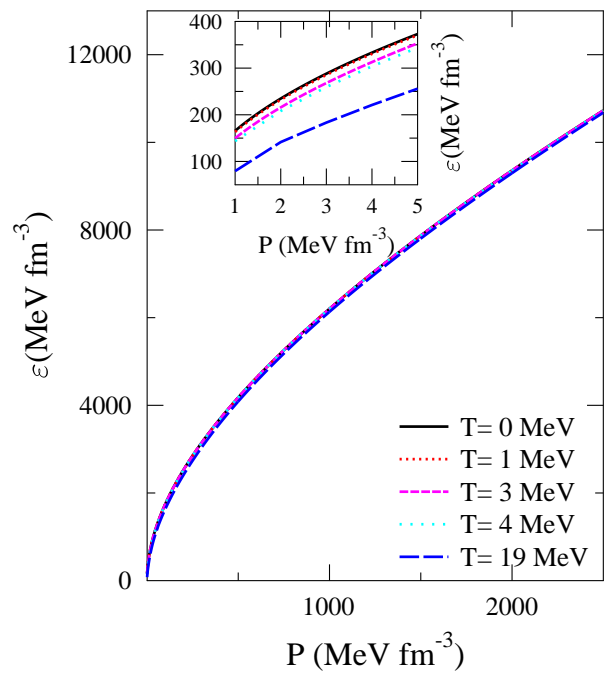, width=8cm}
\caption{(Color online) Equations of state of bosons matter used as input to obtain the mass-radius plots in Fig.~\ref{fig_RMFT}. }
\label{fig_EOSFT}
\end{figure}

By comparing Figs.~\ref{fig_EOST0} and ~\ref{fig_RMT0}, it is obvious that since the role of the  $\frac{1}{2} u_0 n_0^2$ term is to stiffen the EOS at high densities, the maximum mass prediction obtained by taking into account this term in the energy density is lower than that obtained by neglecting this term.  {The reason is clear, i.e., the presence of the  $\frac{1}{2} u_0 n_0^2$ term in this case leads to a linearly increasing binding energy with respect to density, instead of the zero binding energy obtained by Chavanis and Harko~\cite{chavanis}.  The binding energy of boson matter as a function of density for some fixed temperature values is shown in Fig~\ref{fig_BE}. Clearly, the effect of temperature appears only at low densities.} It is also obvious that the difference between the two calculations  {is significant only for the small radius region. Here, we obtain $M_{\rm max} \approx 0.56 M_\odot$, while Chavanis and Harko~\cite{chavanis} predict $M_{\rm max} \approx 0.7 M_\odot$, both with $R_{\rm min} \approx 4.2 ~{\rm km} $.  In this work we note that the Schwarzschild radius for the corresponding maximum mass of the present analysis is $R^{\rm Scd} =2 G M_{\rm max} \lesssim 3.0 ~{\rm km} $, so that {\bf $R^{\rm Scd}/R_{\rm min} \lesssim 0.7 < 0.9$}, which still fulfills the Buchdall limit~\cite{Buchdall}. We also note that the neutron star EOS found in the literature is mostly thermodynamically consistent and it is also well known that the neutron star matter with stiffer EOS predicts a higher maximum mass (see, e.g., Ref. \cite{AS2012} and references therein).

\begin{figure}
\epsfig{figure=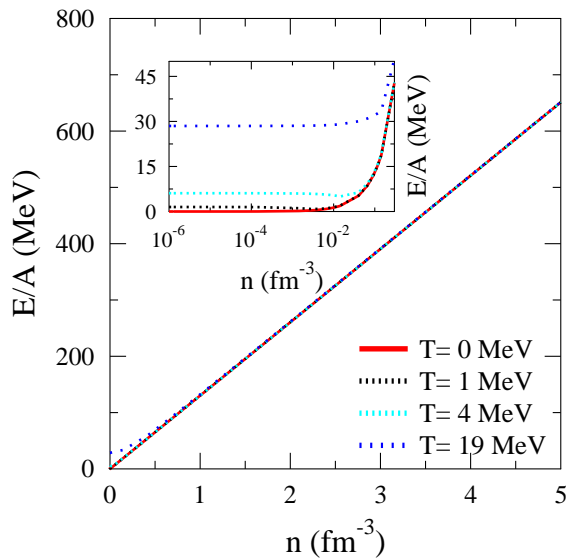, width=7.5cm}
\caption{(Color online) Binding energy as a function of total density and temperature. On the $x$-axes of the inset figure, the logarithmic scale is used. }
\label{fig_BE}
\end{figure}

On the other hand, it is obvious from Fig.~\ref{fig_RMFT} that the temperature variation in a reasonable range, i.e., from 0 to 19 MeV, does not provide a significant effect on the maximum mass of boson star, but for the boson star with relatively large radius, up to the radius of the minimum mass, it can be seen that by increasing the temperature the mass of the corresponding boson star becomes heavier. Note that the mass range of a stable star is from its minimum mass up to its maximum mass, for which on both points ${\partial M(\varepsilon_c) }/{\partial \varepsilon_c}$ = 0~\cite{glendenning}. We can understand this result by observing the corresponding EOS shown in Fig.~\ref{fig_EOSFT}. From this figure it is clear that at high density the EOS does not depend sensitively on the temperature variation, so that the maximum mass predictions are not significantly different, while at low densities, however, the EOS is quite sensitive to the temperature of boson matter. The latter is the main reason for the fact that the mass of boson star with relatively large radius is sensitive to the temperature of boson matter. 

The trend that the EOS of matter does not significantly change due to the variation of temperature is also found in the the case of the isothermal neutron star matter EOS based on Walecka or quark meson coupling models~\cite{PM2010}, as well as in the case of the compact star based on Bag model~\cite{MP2013}. However, the Nambu-Jona-Lasinio model in the case of compact star predicts a relatively different trend at high densities~\cite{MP2013}. This leads to the same mass-radius behavior for boson star and neutron star~\cite{PM2010} due to the temperature variation of each corresponding matter. Finally, we also note that if $T\ge m$, the situation might be different because at such temperature the production of particle-antiparticle pairs cannot be neglected, whereas the the quantum fluctuation and the relativistic effect start to play a significant role \cite{Bilic2000,ABR}.   

In conclusion, by using a simple thermodynamical method we confirm the result obtained by the authors of Ref. \cite{chavanis} that a stable Bose-Einstein condensate star can form. To fulfill the requirement that the  EOS should be thermodynamically consistent, we have obtained a less massive Bose-Einstein condensate star compared to the prediction of Ref.~\cite{chavanis}. We have also found that the maximum mass of boson star is insensitive to the temperature variation. However, the mass of boson star with relatively large radius depends significantly on the temperature of boson matter. The bosons star mass-radius behavior observed in the present work originates from the properties of the interacting boson matter.


This work has been partly supported by the Research-Cluster-Grant-Program 
of the University of Indonesia, under contract No. 1709/H2.R12/HKP.05.00/2014.

\end{document}